\documentclass{jaa}
\usepackage{natbib}
\usepackage[dvipsnames]{xcolor}

\usepackage{graphicx}
\usepackage[T1]{fontenc}

\usepackage{amsmath}	
\usepackage{amssymb}	
\usepackage{textcase}
\usepackage{float}
\usepackage{xspace}
\usepackage[colorlinks,citecolor=blue,urlcolor=blue,bookmarks=false,hypertexnames=true]{hyperref} 

\newcommand{\astro}{\textit{AstroSat}}
\newcommand{\src}{4U 2206+54}
\newcommand{\mnras}{Monthly Notices of the Royal Astronomical Society}
\newcommand{\apj}{The Astrophysical Journal}
\newcommand{\apjl}{The Astrophysical Journal}
\newcommand{\iaucirc}{International Astronomical Union Circulars}
\newcommand{\aap}{Astronomy and Astrophysics}

\newcommand{\apjs}{The Astrophysical Journal Supplement}
\newcommand{\pasj}{Publications of the Astronomical Society of Japan}

\begin{document}\sloppy

\title{Change in spin-down rate and detection of emission line in HMXB \src\ with \astro\ observation}

\author{Chetana Jain\textsuperscript{1,*}, Ajay Yadav\textsuperscript{1} and Rahul Sharma\textsuperscript{2,**}}
\affilOne{\textsuperscript{1}Hansraj College, University of Delhi, Delhi 110007, India.\\}
\affilTwo{\textsuperscript{2}Raman Research Institute, C.V. Raman Avenue, Bangalore 560080 Karnataka, India.}

\twocolumn[{

\maketitle

\corres{*chetanajain11@gmail.com (CJ), **rsharma@rri.res.in (RS)}

\msinfo{July 2022}{August 2022}

\begin{abstract}
This work presents timing and spectral analysis of \src\ using data obtained from LAXPC instrument onboard India's  \astro\ mission. This source was observed with \astro\ in September 2016 and October 2016. We report detection of 5648 (4) s pulsations at MJD 57669 in the latter observation of \src. The pulse profile is sinusoidal and the inherent shape is independent of energy up to 30 keV. The pulse fraction increases with energy from $\sim$0.5\% to $\sim$0.8\%. We report an updated spin down rate of 2.95 (14) $\times$ 10$^{-7}$ s s$^{-1}$. This is about 0.40 times smaller than the previously reported long term value. The energy spectrum is best modelled with an absorbed power-law with high energy exponential cut-off. We have detected presence of broad emission line in \src\ at an energy of 7 keV with equivalent width of $\sim$ 0.4 keV.
\end{abstract}


}]

\doinum{12.3456/s78910-011-012-3}
\artcitid{\#\#\#\#}
\volnum{000}
\year{2022}
\pgrange{1--}
\setcounter{page}{1}
\lp{6}

\section{Introduction}

\src\ was first reported in Uhuru's second catalog and its location was redefined by the Ariel V Sky Survey Instrument \citep{Giacconi72, Villa76}. \src\ is a persistent High Mass X-ray Binary (HMXB) wherein the compact object is accreting matter from a dense wind of an early type star \citep{Steiner84, Negueruela01}. The X-ray light curve shows irregular flaring and $\sim$9.6 d modulation \citep{Negueruela01, Corbet00}. However, from \emph{RXTE} and {Swift} observations, \citet{Corbet07} and \citet{Wang09} have reported modulation at twice the 9.6 d period. The compact object is a strongly magnetized neutron star spinning slowly at a period of $\sim$ 5560 s \citep{Reig09, Torrejon04, Wang09, Finger10}. 

The X–ray spectrum of \src\ is typical of a neutron star accreting from the wind of its companion star \citep{Negueruela01}. The spectrum is known to be well described with a model comprising of absorbed power-law model with exponential cut-off or with a blackbody plus Comptonization \citep{Masetti04, Wang09}. The thermal bremsstrahlung model is also known to describe the \src\ spectrum equally well \citep{Saraswat92}. A possible detection of cyclotron resonant feature has also been reported \citep{Masetti04}. The spectral variations are correlated with the intensity variations \citep{Negueruela01}. The photon index and the hydrogen column density are anti-correlated with hard X-ray luminosity \citep{Masetti04}.

In this paper, the timing and spectral analyses of \src\ have been performed by using the data from Large Area X-ray Proportional Counter (LAXPC) onboard \astro. The paper is organized as follows. Section 2 describes the observation details and the reduction process of the raw data. In Section 3, we discuss the timing analysis of \src. The results from spectral analysis are presented in Section 4. We discuss our results in Section 5.

\section{Observations}

\begin{table}
\caption{Log of \astro-LAXPC observations of \src.}
\centering
\resizebox{\linewidth}{!}{
\begin{tabular}{l l l l l}
\hline
Observation & Observation Time & Mode & Observation & Exposure \\
ID & (yyyy-mm-dd hh:mm:ss) & & span (ks) & (ks)\\
\hline
\\
9000000644 & 2016-09-06 01:33:43 & EA & 88.5 & 46.4 \\
\\
9000000720 & 2016-10-08 16:11:05 & EA & 88.5 & 42.4 \\
\\
\hline
\end{tabular}}
\label{obslog}
\end{table}

\astro\ was launched in September 2015 by Indian Space Research Organization \citep{Agrawal06, Singh14}. The Large Area X-ray Proportional Counter (LAXPC) onboard \astro\ consists of three co-aligned proportional counters (LAXPC10, LAXPC20 and LAXPC30) which cover a broad energy range of 3--80 keV and have a total effective area of 6000 cm$^{2}$ at 15 keV \citep{Yadav16, Agrawal17}. \src\ was observed with \astro-LAXPC twice in 2016. Details of observations are given in Table \ref{obslog}. For the present work, we have not used data from LAXPC30 due to gain issues \citep{Antia17}. We have used event analysis (EA) mode data and all events were extracted from top layer of LAXPC10 and LAXPC20 \citep{Sharma2020}. We have used the LAXPC software (\textsc{LaxpcSoft}: version 3.4.3) to process the level 1 data \footnote{\url{https://www.tifr.res.in/~astrosat\_laxpc/LaxpcSoft.html}}. 

The source and background light curves and spectra were extracted by using the tool \texttt{laxpcl1}. The background products were corrected for gain shift by using \texttt{backshiftv3}. The solar system barycenter correction was performed by using \texttt{as1bary}\footnote{\url{http://astrosat-ssc.iucaa.in/?q=data\_and\_analysis}} tool. We eventually used the tool \texttt{lcmath} to add the background subtracted and barycenter corrected light curves from LAXPC10 and LAXPC20.

\section{Timing Analysis}

\begin{figure*}
\centering
\includegraphics[width=1\columnwidth]{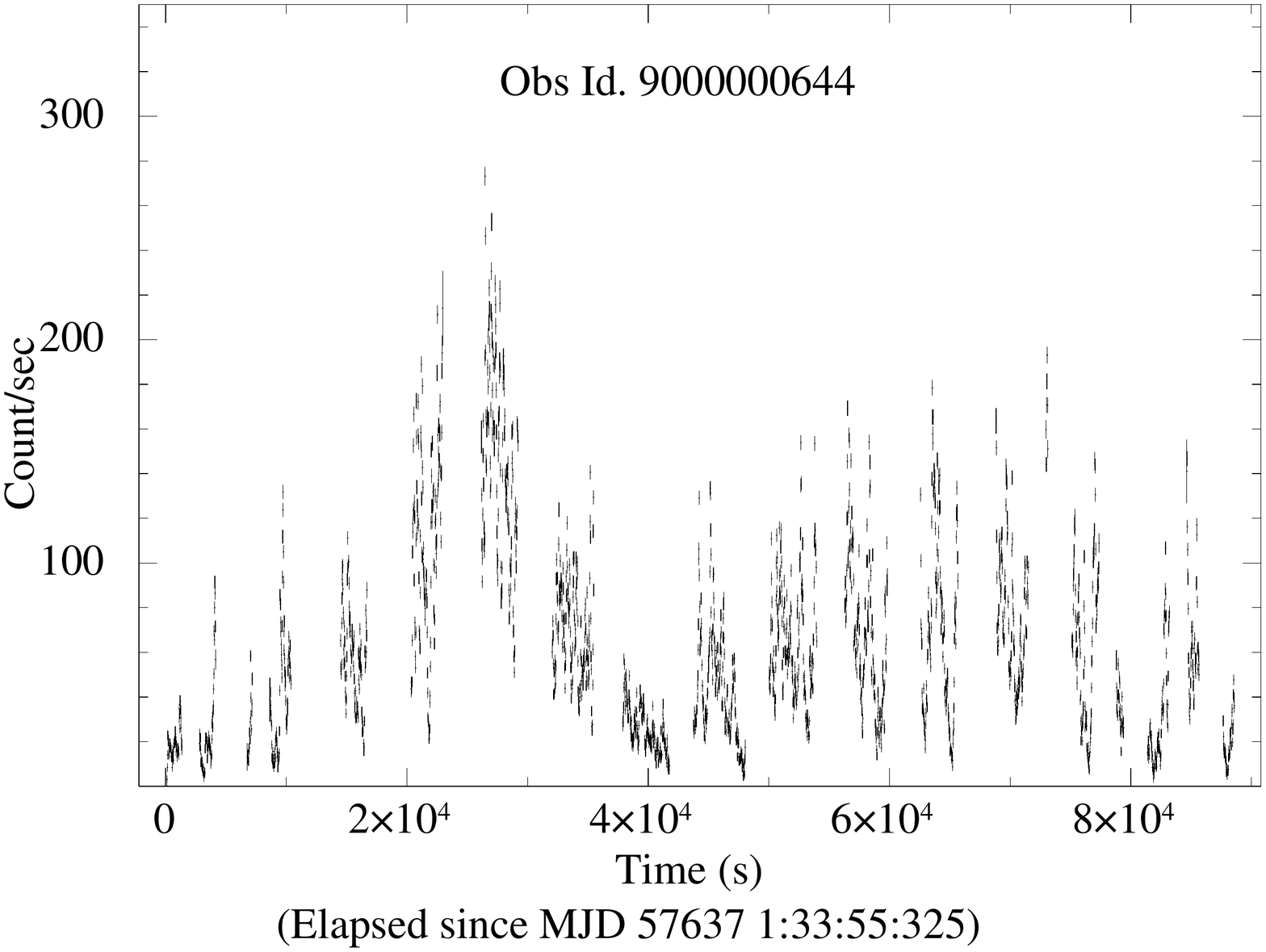}
\includegraphics[width=1\columnwidth]{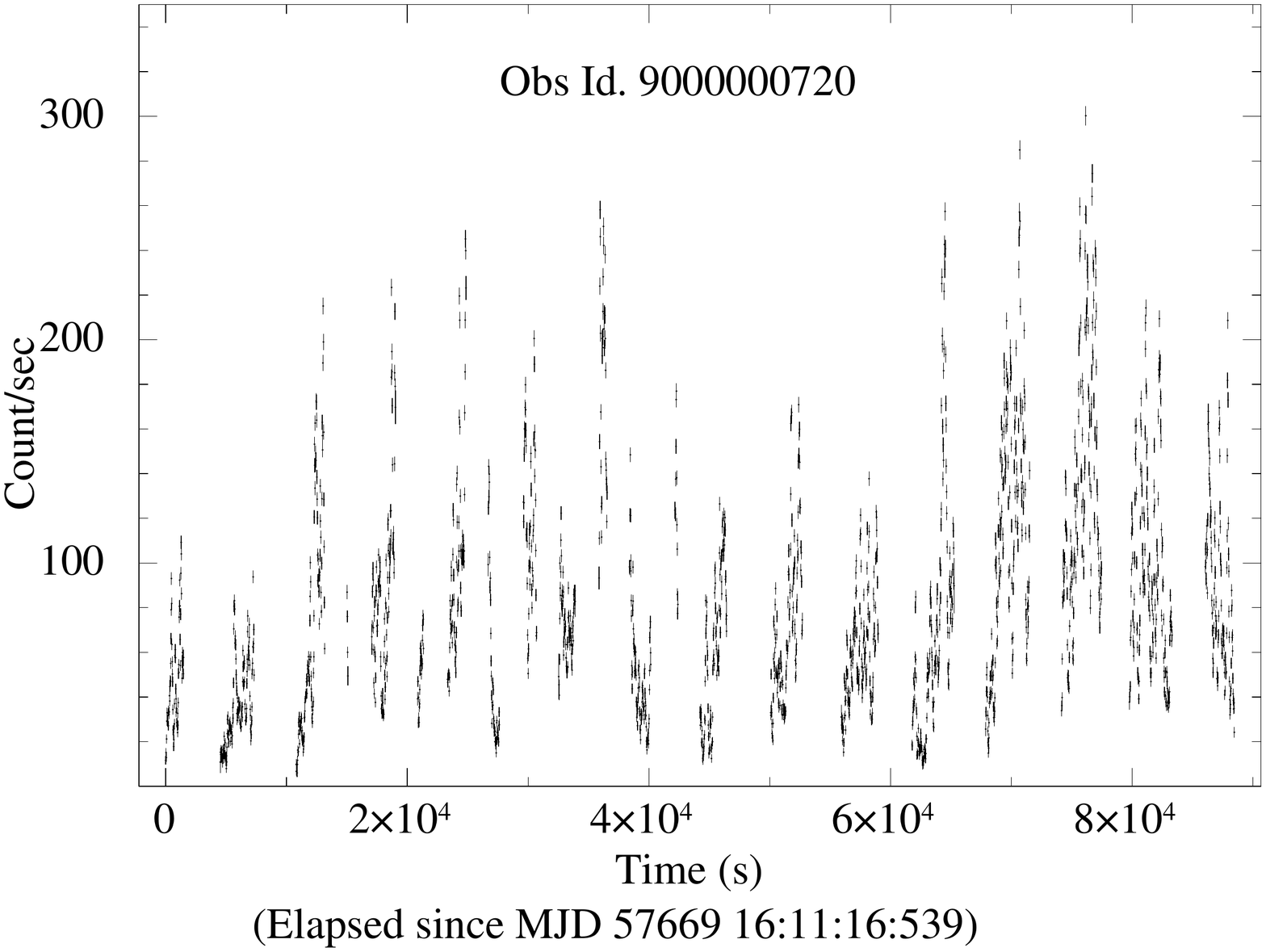}
\caption{\small The 3--20 keV background subtracted and barycenter corrected light curve of \src\ binned with 20 s.}
\label{fig:lc}
\end{figure*}

The 3--20 keV background subtracted and barycenter corrected light curve of \src\ binned with 20 s is shown in Figure \ref{fig:lc}. The X-ray flux is highly variable and exhibits flare-like activity on short timescales similar to that reported by \citet{Negueruela01} for this source and in similar HMXBs by \citet{Yamauchi90} and \citet{Kreykenbohm99}.

\begin{figure}
\centering
\includegraphics[width=1.2\columnwidth]{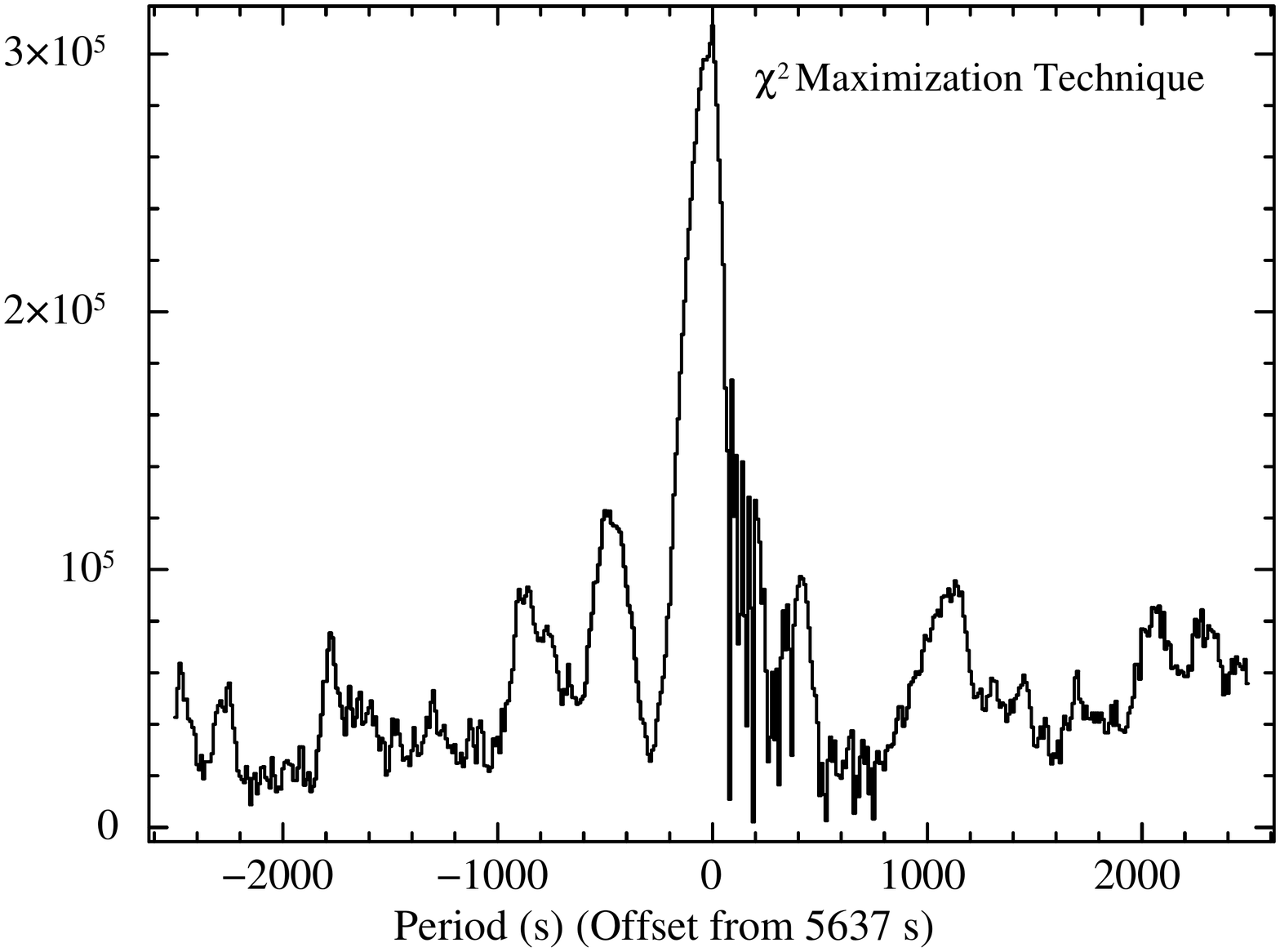}
\includegraphics[width=1.2\columnwidth]{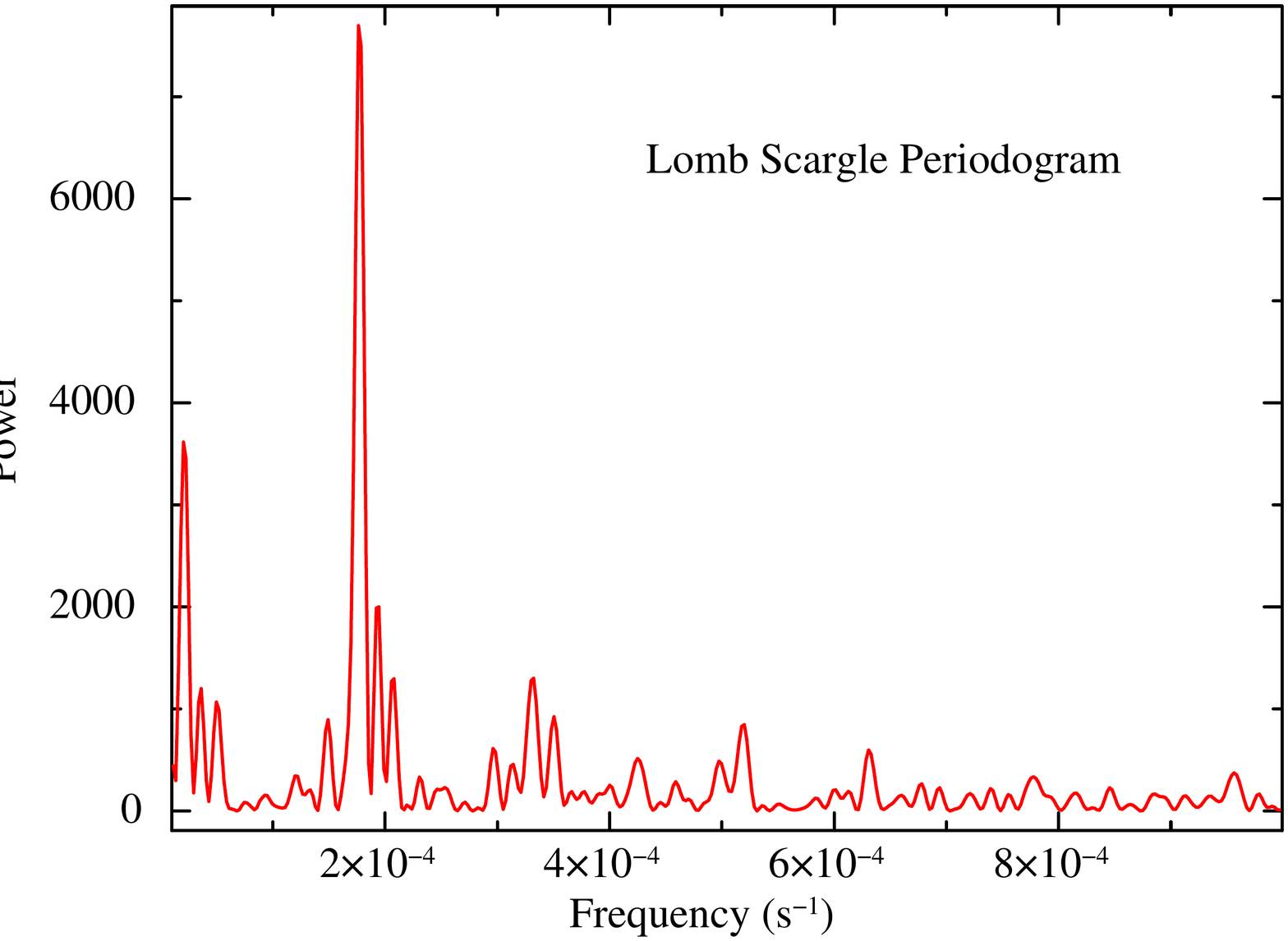}
\caption{\small Plots for spin period determination in \src. \textit{Upper Panel}: $\chi^2$ Maximization Technique. \textit{Lower Panel}: Lomb-Scargle Periodogram}
\label{fig:period}
\end{figure}

We used $\chi^2$ maximization technique to search for spin period in the light curve. Since the spin period is expected to be $\sim$ 5560 s, therefore we folded the time series data over a range of periods (3000--8000 sec) with a resolution of 10 s by using the \texttt{efsearch} tool of XRONOS sub-package of \textsc{FTOOLS} \citep{Blackburn99}. The upper panel of Figure \ref{fig:period} shows the plot of $\chi^2$ values as a function of trial periods. The period corresponding to the maximum $\chi^2$ value was modelled with a Gaussian profile. From the best fit, we obtained a spin period of 5608 s with 1$\sigma$ uncertainty of 4 s. We have detected pulsations only in the October 2016 observation.

We have also used the Lomb-Scargle periodogram method to confirm the periodicity in the unevenly space light curve of \src\ \citep{Scargle82, Press89}. This method has been successfully used to determine periodicity in several other HMXBs \citep{Laycock03, Jain09}. As seen in the lower panel of Figure \ref{fig:period}, a clear peak is present in the periodogram at a frequency of $\sim$ 0.18 mHz. We fitted a Gaussian profile to this peak and obtained a best fit value of the Gaussian center as 0.17706(14) mHz. This corresponds to a pulse period of 5648 (4) s .  

\begin{figure}
\centering
\includegraphics[width=1.0\columnwidth]{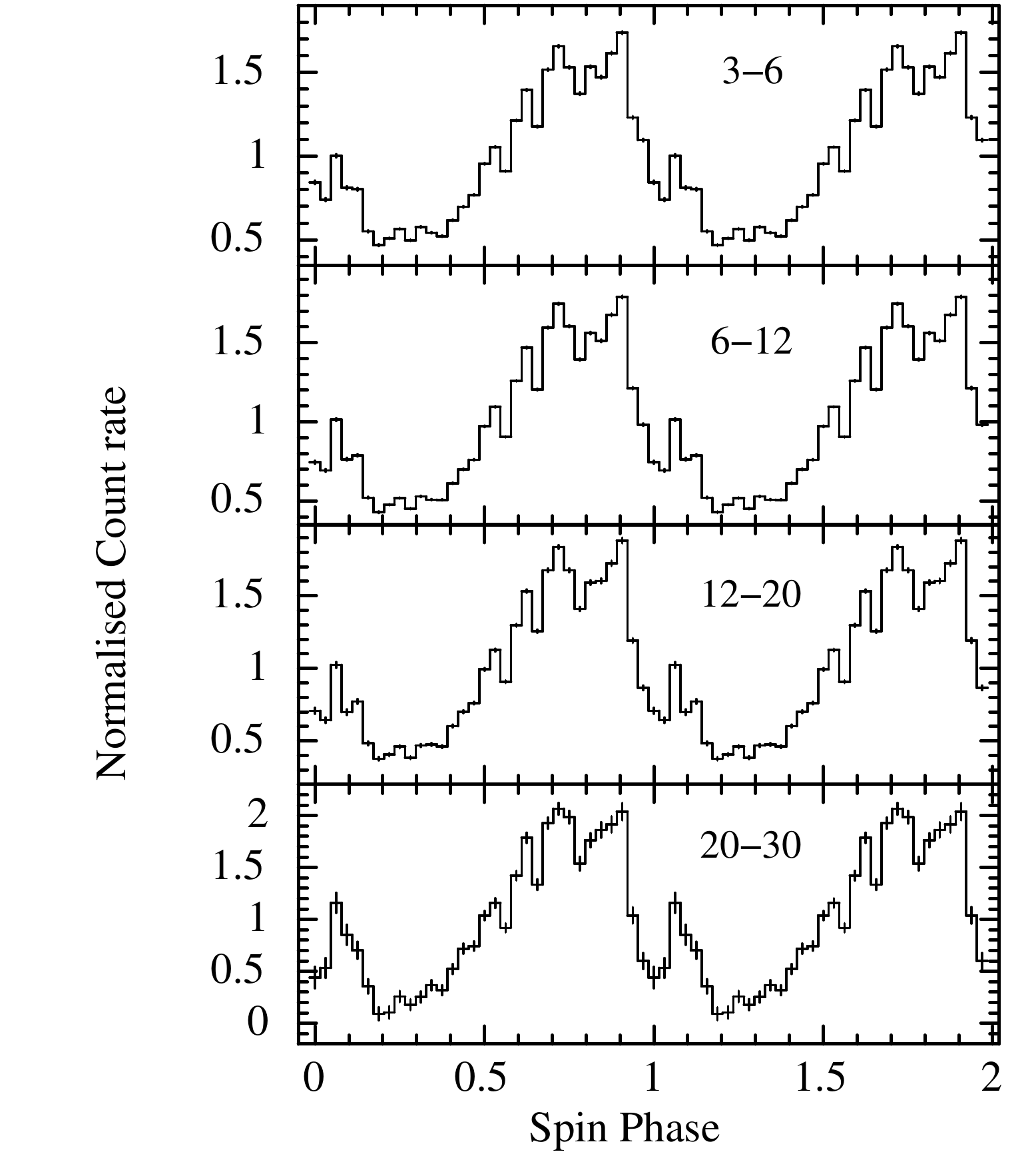}
\caption{\small Energy resolved pulse profile of \src\ in the energy range of 3--6, 6--12, 12--20 and 20--30 keV. Two cycles of spin phase have been shown for clarity.}
\label{fig:efold}
\end{figure}

We created energy resolved pulse profiles using the folding technique with the best period (5648 s) obtained above. Figure ~\ref{fig:efold} shows the energy resolved pulse profiles of \src\ in the energy range of 3--6, 6--12, 12--20 and 20--30 keV, respectively. The inherent sinusoidal shape to the pulse profile was observed to be independent of energy. As a function of energy, the fractional amplitude increased from $\sim 0.5
\%$ to $\sim 0.8\%$. 

\begin{figure}
\centering
\includegraphics[width=1.0\columnwidth]{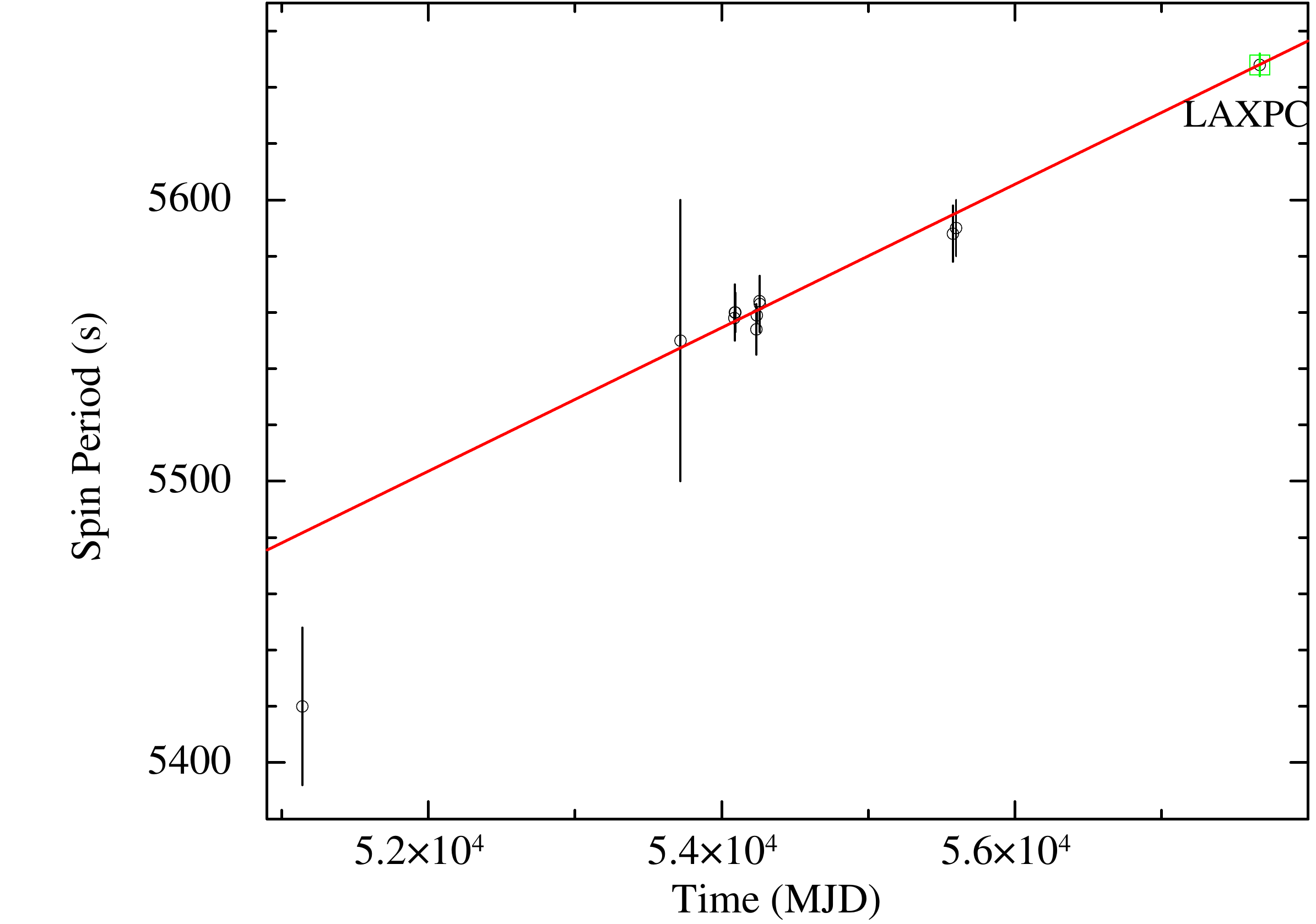}
\caption{\small The spin period evolution of \src\ from 5420 s to 5588 s obtained from \citet{Reig09, Reig12}, \citep{Wang09, Wang13} and \citet{Finger10}. The spin period of 5648 s, determined from this work is shown with green square symbol.}
\label{fig:spintrend}
\end{figure}

From \emph{Suzaku, Beppo}-SAX, \emph{RXTE}, \emph{EXOSAT}, \emph{Integral} and \emph{XMM}-Newton data, \citet{Reig09, Reig12}, \citet{Wang09, Wang13} and \citet{ Finger10} have reported that the spin period of \src\ has increased from about 5525 s to about 5588 s. We have observed a further increase to 5648 s in the spin period with \astro\ observation. Our result gives an updated spin down rate of $\sim 2.95 (14) \times 10^{-7}$ s s$^{-1}$. From numbers fetched from the reported works and the result of the present analysis, Figure ~\ref{fig:spintrend} shows the long term spin evolution in \src.

\section{Spectral Analysis}
During the 2016 \astro\ observations, \src\ was visible above the background up to 30 keV. Therefore, the spectral analysis was done in the energy range 3--30 keV. Due to systematic differences, we have generated spectrum only from LAXPC20. A systematic uncertainty of 1\% was used for the spectral analysis \citep{Antia17, Antia2021}.

\begin{figure*}
\centering
\includegraphics[width=1\columnwidth]{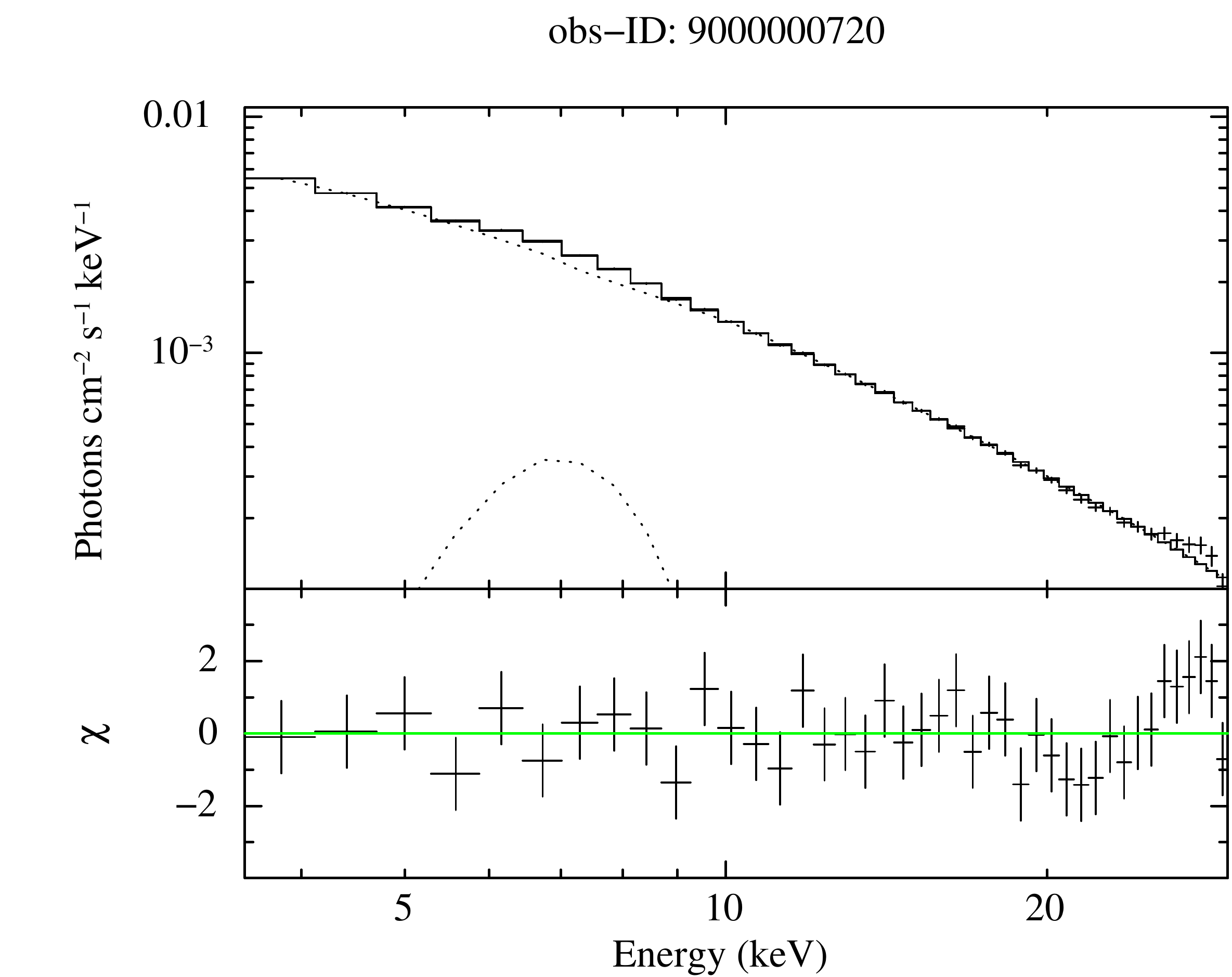}
\includegraphics[width=1\columnwidth]{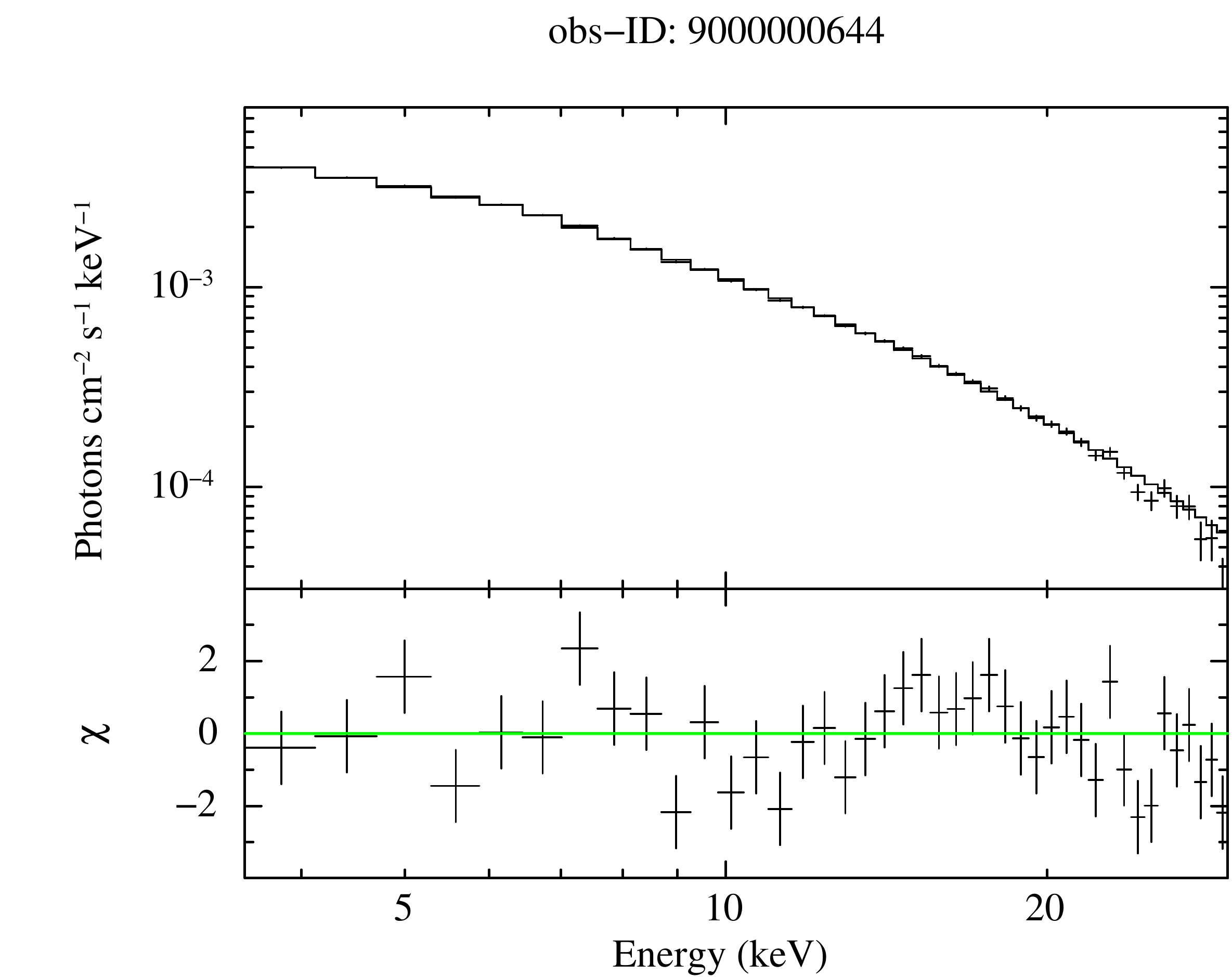}
\caption{\small The 3--30 keV energy spectrum of \src. }
\label{fig:spec}
\end{figure*}

Figure \ref{fig:spec} shows the energy spectrum of \src\ for both the \astro\ observations. We modeled the spectrum using absorbed power-law with high-energy exponential cutoff. The best fit spectral parameters are given in Table \ref{table:spectra}. For observation of October 2016, a broad emission line around 7 keV was also detected with equivalent width of $\sim 0.4$ keV. We did not detect any significant emission feature in the September 2016 observation.

The energy spectrum of \src\ can also be described well with thermal bremsstrahlung model. For both the observations, the broad 6--7 keV emission line was significantly detectable with an equivalent width of $\sim 0.4$ keV. For October 2016 observation, the best fit had a $\chi^2$ of 49 for 36 d.o.f. and $kT_{brem}$ was 18 keV. Addition of a thermal blackbody ($kT_{BB} \sim 2.4$ keV) component improved the fit with a $\chi^2$ of 31 for 34 d.o.f. The $kT_{brem}$ increased to $\sim 25$ keV. For the September 2016 observation, we obtained $kT_{BB}$ of $\sim$ 3.9 keV. 

\begin{table}
\centering
\caption{Best fit spectral parameters of \src\ from LAXPC observations of September 2016 and October 2016. All errors and upper limits reported in this table are at $90\%$ confidence level ($\Delta \chi^2$=2.7).}
\label{table:spectra}
\resizebox{\linewidth}{!}{
\begin{tabular}{l l l l}
\hline
Model & Parameters   & October 2016  & September 2016 \\
\hline
\\
TBabs & $N_H$ ($10^{22}$ cm$^{-2}$) & $5.4 \pm 1.3$ & $1.65^{+1.90}_{-1.65}$ \\
\\
HighEcut & $E_{\rm cutoff}$ (keV) & $10 \pm 2$ & $6.4 \pm 0.4$\\[1ex]
       & $E_{\rm fold}$ (keV) & $26.2_{-4.4}^{+3.8} $ & $10.4^{+1.3}_{-0.96}$ \\
\\      
Powerlaw & $\Gamma$  & $1.69^{+0.08}_{-0.12}$ & $1.04^{+0.15}_{-0.14}$\\[1ex]
        & Norm & $0.070 \pm 0.012$ & $0.017^{+0.006}_{-0.004}$ \\
\\
Gaussian (Fe K) & $E$ (keV)  & $7.0^{pegged}$ & \\[1ex]
        & $\sigma$ (keV) & $1.17^{+0.50}_{-0.45}$ & \\[1ex]
        & Eqw (keV) & $0.422_{-0.351}^{+0.421}$ & \\[1ex]
        & Norm ($10^{-3}$) & $1.14^{+0.75}_{-0.62}$ & \\
\\
Flux$^{a}$ & $F_{3-30 \rm ~keV}$ & $4.6 \times 10^{-10}$ & $3.26 \times 10^{-10}$ \\[1ex]

X-ray Luminosity$^{b}$ & $L_{3-30 \rm ~keV}$ & $3.7 \times 10^{35}$ & $2.6 \times 10^{35}$ \\[1ex]

\\            
          & $\chi^2/{\rm dof}$ & 33.8/35 & 56.7/37 \\

\hline
\multicolumn{4}{l}{$^{a}$Unabsorbed Flux in the units of erg cm$^{-2}$ s$^{-1}$.}\\
\multicolumn{4}{l}{$^{b}$X-ray luminosity in the units of erg cm$^{-2}$.}\\

\end{tabular}}
\end{table}

\section{Discussion}

This work reports results from analysis of \astro-LAXPC data of \src. From the timing analysis we have detected 5648 (4) s pulsations at MJD 57669. Combining our results with known measurements spread across $\sim$18 years, we confirm that this long period neutron star is spinning down at a rate of 2.95(14) $\times$ 10$^{-7}$ s s$^{-1}$. Long spin period and the observed spin evolution rate confirms the magnetar scenario wherein the neutron star had a high magnetic field strength ($\sim$10$^{14}$ G) at birth \citep{Li99, Finger10} which has resulted in spinning down the compact object to current value. The observed spin down rate is about 0.40 times smaller than the reported rate \citep{Finger10, Reig12, Wang13}. It should be noted that since the spectral parameters (except absorption column density) determined in this work are similar to the previously reported numbers, therefore, the reduction in the spin-down rate in \src\ is not associated with a consequential change in the accretion geometry just like in other HMXBs \citep{Baykal06}. In the absence of any clear evidence, it is difficult to comment at this stage on the correlation of the change in spin-down rate with source luminosity. 

From the spectral analysis, we have detected a broad emission line feature at 7 keV. Most of the previous missions have not detected an emission feature except \emph{XMM--Newton} where very weak 6.5 keV fluorescence iron lines have been observed \citep{Reig12}. The authors have reported an absorption column consistent with the interstellar value. This implies very small amount of material around the neutron star and hence a weak detection of emission line. During the \astro\ observation, the absorption column in \src\ was larger than that corresponding to interstellar absorption ($N_H = 0.55 \times 10^{22}$ atoms cm$^{-2}$: \cite{hi4pi}), thereby indicating an optically thick surrounding medium. Hence the detection significance of emission line in the \astro\ observation is much more profound compared to previous works.

\src\ is a highly variable source and its 1-10 keV luminosity is known to vary in the range $\sim$ 10$^{34}$--10$^{36}$ erg s$^{-1}$ \citep{Corbet01, Masetti04, Torrejon04, Wang09}. Even the 3--100 keV luminosity is known to vary by $\sim$ 10$^{35}$--10$^{36}$ erg s$^{-1}$ \citep{Wang13}. We have obtained 3--30 keV luminosity of $L_X \sim 3 \times 10^{35}$ erg cm$^{-2}$ assuming a distance of 2.6 kpc \citep{Blay2006}. A long term X-ray monitoring of \src\ is necessary to study the evolution of spin period of the compact object. Further investigation of the emission lines is also necessary to understand the properties of the circumstellar material surrounding the neutron star. Future observations will also prove to be crucial in understanding the cause for change in the spin-down rate in \src. 

\section*{Acknowledgements}

This work has made use of data from the \astro\ mission of the Indian Space Research Organisation (ISRO), archived at the Indian Space Science Data Centre (ISSDC). We  thank the LAXPC Payload Operation Center (POC) at TIFR, Mumbai for providing necessary software tools. We have also made use of software provided by the High Energy Astrophysics Science Archive Research Center (HEASARC), which is a service of the Astrophysics Science Division at NASA/GSFC. CJ acknowledges the financial grant received from Research and Development Cell of Hansraj College. 


\bibliographystyle{mnras}
\bibliography{refs}

\end{document}